\documentclass[runningheads]{llncs}

\usepackage[T1]{fontenc}
\usepackage{graphicx}
\usepackage{multirow}
\usepackage{tablefootnote}
\usepackage{hyperref}
\usepackage{tabularx}
\usepackage{textcomp}
\usepackage{subcaption}
\usepackage{setspace}
\usepackage{color}

\begin{document}

\title{Unified Shared Memory in OpenMP: Implementation, Programmability, and Performance on Intel Accelerators}

\titlerunning{USM in OpenMP on Intel Accelerators}
%
\author{Harald Servat\inst{1}\orcidID{0000-0002-0144-7934} \and Fran\c{c}ois Dugast\inst{1}\orcidID{0009-0009-2994-0358} \and Alejandro Duran\inst{1} \and Abhinav Gaba\inst{1} \and Rakesh Krishnaiyer\inst{1}}

\authorrunning{Harald Servat et al.}

\institute{Intel Corporation\\
\email{name.surname@intel.com}}

\maketitle

\begin{abstract}

OpenMP 5.0 introduced the Unified Shared Memory (USM) feature through the \texttt{requires} directive. The feature simplifies the adoption of the OpenMP programming model by providing a unique and common address space between the accelerators and the host and allowing the access (dereference) of the same memory address on different devices, thus avoiding the burden of explicit data transfers to maintain the consistency between the address spaces. Hence, the feature eases quick prototyping and porting of applications to OpenMP with accelerators.

In this paper, we introduce the Intel implementation for USM. We briefly discuss its implementation in the software stack (OS kernel, compiler, and runtime), then assess its adoption complexity in existing HPC applications using OpenMP for accelerators, and, finally, evaluate the performance of these applications when adopting USM on an Intel Battlemage GPU. USM is not expected to grant performance uplifts to already optimized applications with explicit, granular data-motion control and our results show an overhead with a geometric mean below $1.2x$ ($1.03x$ seems achievable with further optimizations). Yet, in this paper we show there exist applications that benefit from this feature, making it attractive even for already ported applications.
\end{abstract}

\section{Introduction}

Nowadays, General Purpose GPUs (GPGPUs) are ubiquitous in High-Performance Computing (HPC) and data centers because they offer significant computational potential over CPUs. OpenMP* has been offering offload capabilities to these devices for more than a decade since version 4.0, which consolidated the offload features from the first Technical Report (TR1)~\cite{OpenMP40TR1}. However, one potential obstacle that hinders the adoption of this programming model is the fact that accelerators and the host have different data environments. When a host variable is mapped to an accelerator, unless the host and device can access the same storage locations, a corresponding variable needs to be allocated in the device's memory, and if host and device cannot access the same storage locations, copy operations are required to make original and corresponding variables consistent.

OpenMP 5.0 added the \texttt{requires} directive to support applications that require implementation-specific features. Among those features, we find \texttt{unified\_shared\_memory} by which the implementation guarantees that all devices use a unified address space and that storage locations in memory are accessible to all devices. 
This allows the implementation to skip allocating any device memory for corresponding variables, and instead pass through the original variable's address to the device, bypassing the need for explicit data transfers in most cases.
Consequently, this feature eases the adoption of the programming model. Unified Shared Memory (USM) has been actively adopted in HPC frameworks~\cite{heroux2005overview,reyes2020sycl,10.1145/3648115.3648118,10.1145/3648115.3648131} and is still evolving across platforms from the major GPU vendors (including NVIDIA, AMD and Intel).
The only requirement imposed by the \texttt{unified\_shared\_memory} directive is that it has to be present in every compilation unit containing device constructs, device procedures or \texttt{declare target} directives. This simplicity makes it an attractive feature for developers prototyping their applications in OpenMP with accelerators because it removes the burden of maintaining data consistency across devices and allows them to focus on the computational part.

Regarding performance, a common expectation is that the simplicity offered by this feature comes at the price of overhead. In this paper, we evaluate the performance of several HPC-related benchmarks and applications from different fields already using OpenMP for accelerators and explore the performance variation when adopting \texttt{unified\_shared\_memory}. We want to stress that this might not be the most common way to explore this directive because one expects it to be used in the early stages of development and because the development at later stages might have already optimized explicit data transfers, and thus, it seems unlikely that the directive benefits the application. However, we show that it is worth exploring the usage of the aforementioned directive even on these applications because it can simplify the code and potentially improve performance.

The main contributions of this paper are:
\begin{itemize}
    \item \textbf{Intel USM implementation:} Description of the full software stack (OS kernel, userland driver, compiler and runtime) for OpenMP's \texttt{unified\_shared\_memory},
    \item \textbf{Performance evaluation:} benchmarking the impact of adopting USM on existing HPC OpenMP offload applications, challenging the assumption it always adds overhead, and,
    \item \textbf{Future directions:} discussion of several topics of interest related to USM, including, but not limited to: OpenMP extensions, performance analysis, and potential suggestions for compilers.
\end{itemize}

\section{Related Work}
\label{sec:RelatedWork}

Managing separate host and accelerator memory pools is error-prone and laborious. Some vendors have addressed this through unified memory abstractions, as in the case of HSA~\cite{HSA}, a multi-company standard led by AMD, ARM, and Samsung, that introduced shared virtual memory and system-wide atomics for heterogeneous platforms. Later, OpenCL 2.0~\cite{OpenCL20_URL} included support for Shared Virtual Memory (SVM) that has to be managed through SVM-specific routines (e.g. \texttt{clSVMAlloc()} and \texttt{clSVMFree()}). SYCL 2020~\cite{SYCL2020_URL}, which was inspired by OpenCL, added support for Unified Shared Memory (USM). Both SVM and USM provide a unified address space for host and device, but the latter extends SVM with full support for pointer arithmetic and fine-grained control over ownership and accessibility of memory allocations~\cite{10.1145/3388333.3388644}. Finally, there have been proposals for improving OpenMP current allocation and memory management features~\cite{10.1007/978-3-030-85262-7_13}. The authors of this work believe that an all-or-nothing approach on unified memory is not sufficient, and they propose extending OpenMP memory allocation interfaces to include a memory space abstraction with device awareness and a concept of shared memory.

In the GPU vendor realm, NVIDIA's Managed Memory, available since NVIDIA* CUDA 6~\cite{harris2013unified}, automates data movement for regions allocated via \texttt{cudaMallocManaged()}. Early limitations—migration only at kernel boundaries, managed memory capped at GPU physical memory size, inability to work on stack arrays, and unnecessary host-to-device copies—made it difficult to match manual data management performance~\cite{landaverde2014investigation,li2015evaluation}. CUDA 8 and the NVIDIA* Pascal GPU architecture~\cite{harris2016cuda} introduced NVIDIA Unified Memory, which removed these restrictions, enabling concurrent CPU/GPU access, expanded managed memory, and system-wide atomics. On top of that, CUDA Fortran~\cite{CUDAFortran_URL} allows tagging individual variables for placement in unified memory via attributes, or alternatively using compiler flags to place all variables in unified memory. OpenACC*~\cite{OpenACCUnifiedMemory_URL} only supports compiler flags to place all host data in unified memory and does not support data prefetching, although CUDA prefetching functions can be called directly. Regarding AMD, AMD* ROCm also supports Unified Shared Memory since their version 6.0 for their AMD* Instinct starting with the MI200 series GPUs~\cite{AMD_ROCM_URL}. AMD relies on the Heterogeneous-Compute Interface for Portability (HIP) and Xnack to support GPU page faulting on system or managed allocations.

\section{Intel's Implementation of Unified Shared Memory}
\label{sec:Implementation}

The Intel software stack around USM for OpenMP involves four components: 1) the Linux kernel module (KMD), or Xe driver; 2) the userland driver (UMD); 3) the C/C++ and Fortran compilers with OpenMP support; and 4) the OpenMP runtime. We briefly describe the USM implementation in each component.

\subsection{Implementation in the Linux Xe driver}
\label{sec:implementation-kmd}

The USM feature is implemented in the Linux Xe driver~\cite{DRM_XE_Intel_GFX_Driver_URL} using a demand-paging model. Rather than requiring applications to explicitly manage GPU memory allocations, the GPU virtual address space is mapped as a mirror of the CPU process address space. GPU page tables are left unpopulated at allocation time; when the GPU accesses an unmapped address, the hardware raises a page fault that is forwarded to the driver via the Graphics \textmu-Controller (GuC) firmware. The driver services the fault by querying the state of the CPU page tables for the faulted range, computing the appropriate GPU page table entries, and installing them before signaling the firmware to resume the stalled engine. This lazy population model avoids the cost of pre-mapping the entire address space and allows the GPU to naturally follow the working set of the application.

The driver exploits the availability of on-card high-bandwidth memory (VRAM) by transparently migrating data pages from system memory to VRAM at fault time. When a GPU page fault is serviced and conditions are met, the driver allocates VRAM storage and uses the GPU's copy engine to transfer the data before installing GPU page table entries pointing to VRAM. To achieve this, VRAM is registered with the OS as a zone of device-private memory, giving each VRAM page frame a first-class kernel page descriptor. This allows DRM GPU SVM helpers built on the Heterogeneous Memory Management (HMM) subsystem to treat VRAM and system memory pages uniformly during migration. When available, Transparent Huge Pages (THP) allow 2 MB migration granularity to amortize fault-handling overhead over contiguous regions, which is particularly beneficial for the large, regular array accesses typical of HPC workloads.

Coherence between the CPU and GPU views of memory is maintained through the kernel MMU notifier mechanism. Whenever the OS modifies the CPU page table (due to reclaims, migrations, compactions, among others) for a range that has an active GPU mapping, the driver is notified synchronously. It invalidates the corresponding GPU page table entries and issues a TLB invalidation request to the GPU through the firmware, waiting for acknowledgement before allowing the CPU-side operation to proceed. This strict ordering guarantees that the GPU cannot use a stale mapping after the underlying physical page has been reassigned. When a CPU thread accesses a page that currently resides in VRAM, or when the OS reclaims VRAM capacity under memory pressure, the data is migrated back to system memory via the GPU copy engine; any GPU mapping for the affected pages is invalidated and will be re-established on the next GPU access. The result is a coherent, shared address space in which data placement is managed automatically by the driver.

\subsection{Implementation in the UMD}
\label{sec:implementation-umd}

The User Mode Driver (UMD) abstracts the KMD device through APIs such as Intel\textsuperscript{\textregistered} oneAPI Level Zero~\cite{LevelZero_URL} or OpenCL. It consists of the NEO libraries, the Intel\textsuperscript{\textregistered} Graphics Memory Management Library (GMM), and the Intel\textsuperscript{\textregistered} Graphics Compiler for OpenCL\textsuperscript{\texttrademark} (IGC). Its main responsibilities are managing device memory, JIT-compiling kernels, and providing device queues and synchronization primitives.

Shared System USM support in the Xe driver allows the GPU to directly access CPU-allocated memory — including ordinary \texttt{malloc()} allocations — without explicit copies or driver-managed allocations. This requires hardware support for CPU address mirroring, recoverable page faults, and a GPU address space large enough to cover the full CPU virtual address range. When these conditions are met, the driver maps the system address space into the GPU's page table at virtual memory (VM) creation time. Memory advisory hints (\textit{e.g.} preferred location or atomic access mode) and prefetch operations then guide data migration, ensuring efficient data moves between CPU/GPU on demand.

\subsection{Compiler Support for Unified Shared Memory}
\label{sec:implementation-compiler}

For this study, we use the Intel\textsuperscript{\textregistered} oneAPI Toolkit version 2026.0 compilers~\cite{OneAPIToolkit2026_URL} that support OpenMP offloading to Intel GPUs (either discrete as Battlemage, Ponte Vecchio, or integrated such as PantherLake, \textit{etc.}) using the Level Zero libomptarget plugin. The C/C++/Fortran optimizing compiler is based on the LLVM framework and supports more than 90\% and over 40\% of OpenMP 5.x and 6.x features, respectively – including advanced \texttt{omp target} related features.

When the programmer writes \texttt{\#pragma omp requires unified\_shared\_memory}, 
the compiler makes adjustments to IR generation to support USM semantics.

\begin{itemize}
\item Sets the \texttt{OMP\_REQ\_UNIFIED\_SHARED\_MEMORY} bit in its
\texttt{RequiresFlags}. The value of this flag is embedded in the
generated IR and consulted by the runtime to determine whether USM needs to be enabled.

\item For \texttt{declare target} global variables marked
\texttt{enter}, the compiler suppresses the creation of device-side offload
entries on the target device, since USM makes such globals directly accessible without explicit mapping.

\item When registering device global variables, the normal path is skipped and an indirection pointer 
is created instead, and an entry for it is created in the device offload table, indicating that the pointer 
address must be adjusted to point to the host address.
\end{itemize}

\subsection{Runtime Support for Unified Shared Memory}
\label{sec:implementation-runtime}

When the runtime is initialized, if it detects that the \texttt{OMP\_REQ\_UNIFIED\_SHARED\_MEMORY} flag
is set for any module in the binary, it performs a number of actions:
\begin{itemize}
\item It tells the Level Zero plugin to enable USM support for the device if available. The plugin will verify 
    that all available devices support USM by checking the \texttt{ZE\_MEMORY\_ACCESS\_CAP\_FLAG\_CONCURRENT} 
    bit in the \texttt{sharedSystemAllocCapabilities} field of the device memory access properties.
\item It initializes the indirect pointers of global variables for each module to point to the host addresses of 
      the variables, so that device code can access them directly without explicit mapping.
\end{itemize}

While OpenMP only guarantees the map operations to not result in any corresponding allocation under \texttt{requires self\_maps}, in practice, Intel's implementation (which aligns with clang's) decides to skip memory allocation for corresponding variables under \texttt{requires unified\_shared\_memory} too, except for \texttt{map(close)}. This allows the flexibility of letting users skip any explicit data motion to let KMD/UMD automatically migrate data as mentioned in Sections~\ref{sec:implementation-kmd} and \ref{sec:implementation-umd}, while also granting some control to \textit{force} allocation in cases where it may be beneficial.

Once USM is active, the following changes to the runtime behavior take place:
\begin{itemize}
\item The runtime does not need to perform explicit data movement for mapped variables, if corresponding allocation for them was skipped, as discussed above,
since they are directly accessible from the device, eliminating unnecessary copies.
\item The runtime can optimize pointer accessibility queries by treating all pointers as accessible 
      when USM is enabled, which can reduce overhead in some cases.
\item The runtime can allocate private kernel arguments in host-accessible USM memory, which can avoid 
      device-side allocation for arguments.
\end{itemize}

\section{Evaluation}
\label{sec:Evaluation}

For the evaluation, we use a system consisting of an Intel\textsuperscript{\textregistered} Core\textsuperscript{\texttrademark} Ultra 5 245K~\cite{CoreUltra5_URL} (ArrowLake) and an Intel\textsuperscript{\textregistered} Arc\textsuperscript{\texttrademark} Pro B70 GPU~\cite{ArcProB70_URL} (Battlemage), each with 32~GBytes of memory. Battlemage is the first Intel GPU to support HW page-faulting. The system runs Ubuntu 24.04 with the vanilla Linux kernel 7.1rc1 and the Intel\textsuperscript{\textregistered} Graphics Compute Runtime for oneAPI Level Zero version 26.18.38308.1~\cite{GraphicsComputeRuntime_URL} along with oneAPI Level Zero Loader version 1.28.2~\cite{OneAPILevelZeroLoader_URL} and IGC version 2.34.4~\cite{IGC_URL}. Applications are compiled with Intel OneAPI Toolkit version 2026.0. To evaluate the performance, we have set the CPU and the GPU to their maximum frequencies (3.7 and 2.8~GHz, respectively) but both devices may operate at lower frequencies depending on the thermal and power conditions. We have reduced the OS verbosity by setting the DRM debug parameter\footnote{Accessible through \texttt{/sys/module/drm/parameters/debug}} to 0 and keeping the GuC log verbosity at the default value (1)\footnote{The contents of this variable can only be altered when loading the xe module as in \texttt{modprobe xe guc\_log\_level=1}.}.

We evaluate the impact of USM in several benchmarks that already implement OpenMP offload with either explicit or implicit data transfers. Table~\ref{tbl:ListApps} lists the evaluated benchmarks along with their respective fields and/or domains of application, and also the main programming language. Note that HipFT uses the \texttt{do concurrent} constructs, which in the Intel\textsuperscript{\textregistered} Fortran Compiler can automatically offload using compiler-generated OpenMP target constructs under a command-line flag\footnote{-fopenmp-target-do-concurrent}. As of writing this document, \texttt{declare target} on globals does not work on the Intel Fortran Compiler and a fix is being worked on, so we manually removed these constructs from the Fortran sources too. Our evaluation focuses on two different aspects: programmability and performance.

\begin{table}[t]
\small
\begin{minipage}{\textwidth}
\caption{List and description of the benchmarks explored in this paper.}
\label{tbl:ListApps}
\begin{tabularx}{\textwidth}{l l l l}
\textbf{Bench name} & \textbf{Suite} & \textbf{Field / Domain} & \textbf{Language} \\
\hline
403.stencil & \multirow{7}{6em}{\rotatebox[origin=c]{90}{SPECaccel\textsuperscript{\textregistered}2023} \cite{SPECACCEL}~\cite{SPECACCEL_URL} } &  Thermodynamics & C \\
404.lbm & &  Fluid dynamics, Lattice Boltzmann & C \\
450.md & &  Molecular dynamics & Fortran \\
452.ep & &  Embarrassingly parallel & C \\
459.miniGhost & &  Finite difference & Fortran \\
460.ilbdc & &  Fluid mechanics & Fortran \\
463.swim & &  Weather & Fortran \\
\hline
miniFE\footnote{\url{https://github.com/ORNL/HeCBench/commit/9c6a275621b626b314d91a5c962f926ad1e20fec}} & \multirow{2}{6em}{Mantevo~\cite{MantevoSuite}} & Implicit finite element & C++ \\
miniMD\footnote{\url{https://github.com/Mantevo/miniMD/commit/2662065eb2b2264a673bda780a23d296f05a2c87}} & & Molecular dynamics & C++ \\
\hline
HipFT~\cite{HipFT_2025}\footnote{\url{https://github.com/predsci/HipFT/commit/fba3b04ec9b0f48e35c160c12424ea208753e340}} & - & Solar physics & Fortran (do concurrent)~\cite{HipFT_2024_portability} \\
\hline
LULESH~\cite{Lulesh}\footnote{\url{https://github.com/ORNL/HeCBench/commit/9c6a275621b626b314d91a5c962f926ad1e20fec}} & - & 3D Lagrangian hydrodynamics & C++ \\
\hline
\end{tabularx}
\end{minipage}
\end{table}

\subsection{Programmability}

In this section, we evaluate the impact of using USM in the preceding codes in terms of programmability. Our analysis covers two scenarios: 1) the compiler supports the automatic USM approach (\textit{i.e.} using \texttt{\#pragma omp requires unified\_shared\_memory}) and 2) the compiler does not support it (\textit{i.e.} needing manual intervention to use USM). In the remainder of this document, we will refer to the former as \textit{automatic-USM} and to the latter as \textit{manual-USM}.  

This differentiation helps us evaluate the impact on a code to be run on a system already supporting USM but at a stage where the compiler does not yet support the aforementioned OpenMP directive -- or a situation where a developer prefers to stay away from the directive. When the compiler supports the directive, getting access to USM is straightforward because the developer only has to add the directive on every compilation unit. However, when the compiler does not support it, the developer can partially mimic its behavior through the following steps (although there may be situations where this approach may not work):
\begin{enumerate}
\item Remove or avoid \texttt{target enter|exit data} constructs,
\item Remove or avoid \texttt{target update} constructs,
\item Remove or avoid \texttt{declare target()} constructs for variables, and
\item Replace \texttt{map()} with \texttt{is\_device\_ptr()} or \texttt{has\_device\_addr()} clauses on target constructs, as appropriate.
\end{enumerate}
Note that the last step only refers to explicit data transfers. The compiler optimization reports identify implicit data transfers, and if any exist, these must be explicitly added in the source code honoring step 4.

\begin{table}[t]
\small
\caption{Impact on the application Lines of Code (LoC) to support USM based on the availability and applicability of \texttt{\#pragma omp requires unified\_shared\_memory}.}
\label{tbl:ListApps_w_USM_LOC}
\begin{tabularx}{\textwidth}{p{0.18\textwidth} >{\centering\arraybackslash}p{0.10\textwidth} >{\centering\arraybackslash}p{0.20\textwidth} >{\centering\arraybackslash}p{0.10\textwidth} >{\centering\arraybackslash}p{0.10\textwidth} >{\centering\arraybackslash}p{0.20\textwidth}}
\textbf{Bench name} & \textbf{\# files} & \textbf{\# OMP stmts} & \multicolumn{3}{c}{\textbf{LoC}} \\
& & & \multicolumn{2}{c}{\textbf{automatic-USM}} & \multicolumn{1}{c}{\textbf{manual-USM}} \\
& & & \multicolumn{1}{c}{\textit{Blind}} & \multicolumn{1}{c}{\textit{Selective}} & \\
\hline
403.stencil & 11 & 3 & +5 & +2 & +3 \\
404.lbm & 6 & 11 & +2 & +2 & +18 \\
450.md & 21 & 83 & +34 & +14 & +32 \\
452.ep & 12 & 13 & +5 & +1 & +12 \\
459.miniGhost & 29 & 199 & +30 & +15 & +65 \\
460.ilbdc & 6 & 19 & +5 & +4 & +13 \\
463.swim & 1 & 164 & +3 & +3 & +72 \\
\hline
miniFE & 23 & 24 & +3 & +1 & +60 \\
miniMD & 26 & 114 & +11 & +9 & +59 \\
\hline
HipFT & 1 & 76 & +22 & +1 & +213 \\
\hline
LULESH & 5 & 52 & +4 & +4 & +82 \\
\hline
\end{tabularx}
\end{table}

At this point, we want to stress that \texttt{\#pragma omp requires unified\_shared\_memory} is only needed in compilation units with OpenMP offload directives. However, selectively adding this directive is error prone and time-consuming, especially if we consider included (or header) files or codes under heavy active development. Hence, we consider that users might consider the blind (and simple) approach of adding the directive to every compilation unit.
Table~\ref{tbl:ListApps_w_USM_LOC} illustrates the number of source files (including headers), the number of OpenMP statements and the three rightmost columns show the lines of code (LoC) added when comparing the \textit{automatic-USM} (using either a blind or selective approach) and \textit{manual-USM} versions against the original.
The numbers in the Table show that \textit{automatic-USM} reduces the amount of changes with a few exceptions, especially when selectively applying the directive to those compilation units with OpenMP offload directives, which aligns with our expectations. Still, a blind addition to every compile unit keeps the LoC added low, except for Fortran applications (450.md, 459.miniGhost and HipFT) where these applications either have more files or have some files with Fortran subroutines which are not encapsulated into a module, and consequently, every subroutine requires the construct.

\subsection{Performance}

\begin{figure}[t]
	\centering
    \begin{subfigure}{0.485\textwidth}
		\centering
        \includegraphics[scale=0.315]{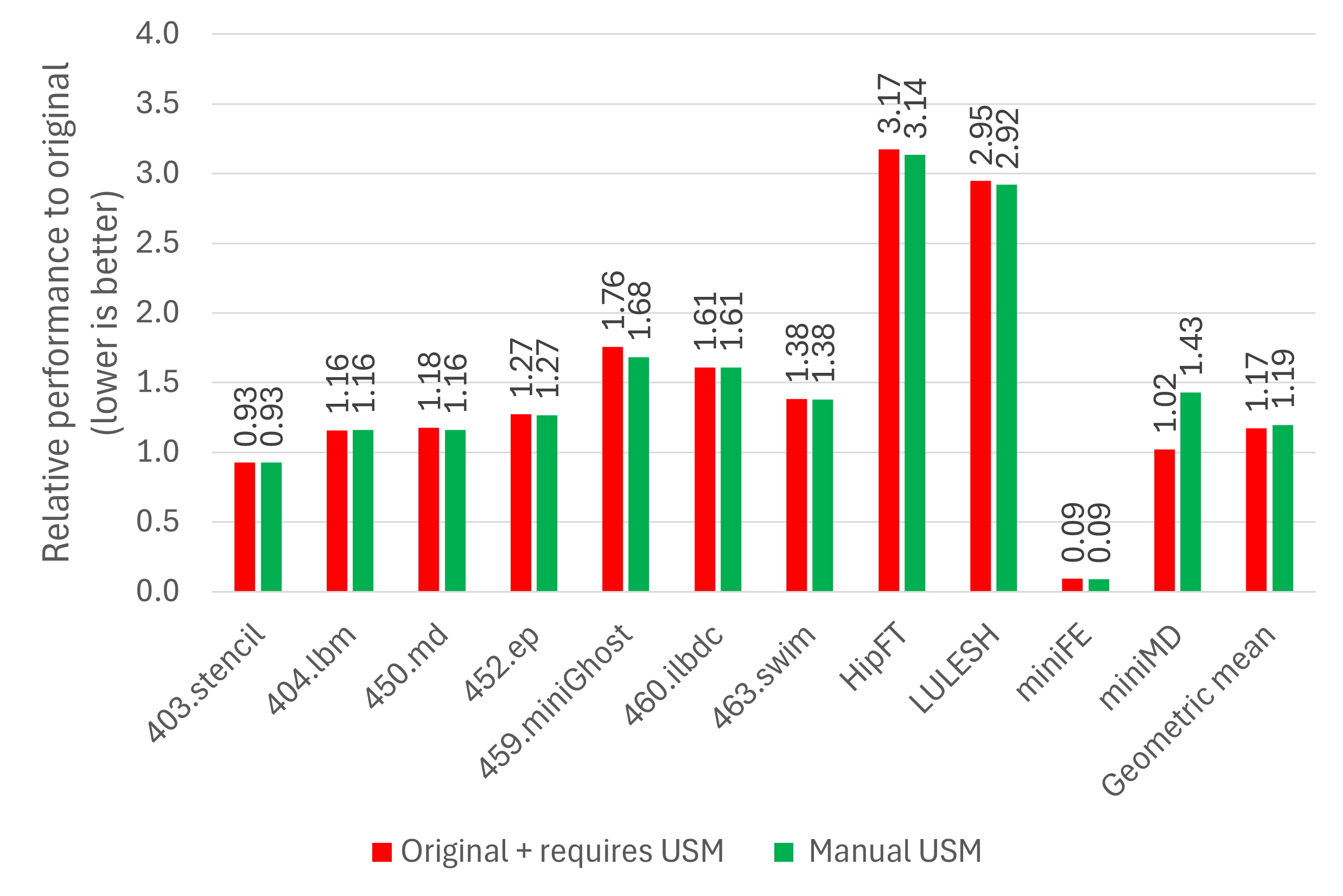}
        \caption{Relative performance of the USM versions when compared against the original version using explicit data transfers.}
        \label{fig:RelativePerformanceToOriginalVersion}
    \end{subfigure}
	~
    \begin{subfigure}{0.485\textwidth}
        \includegraphics[scale=0.315]{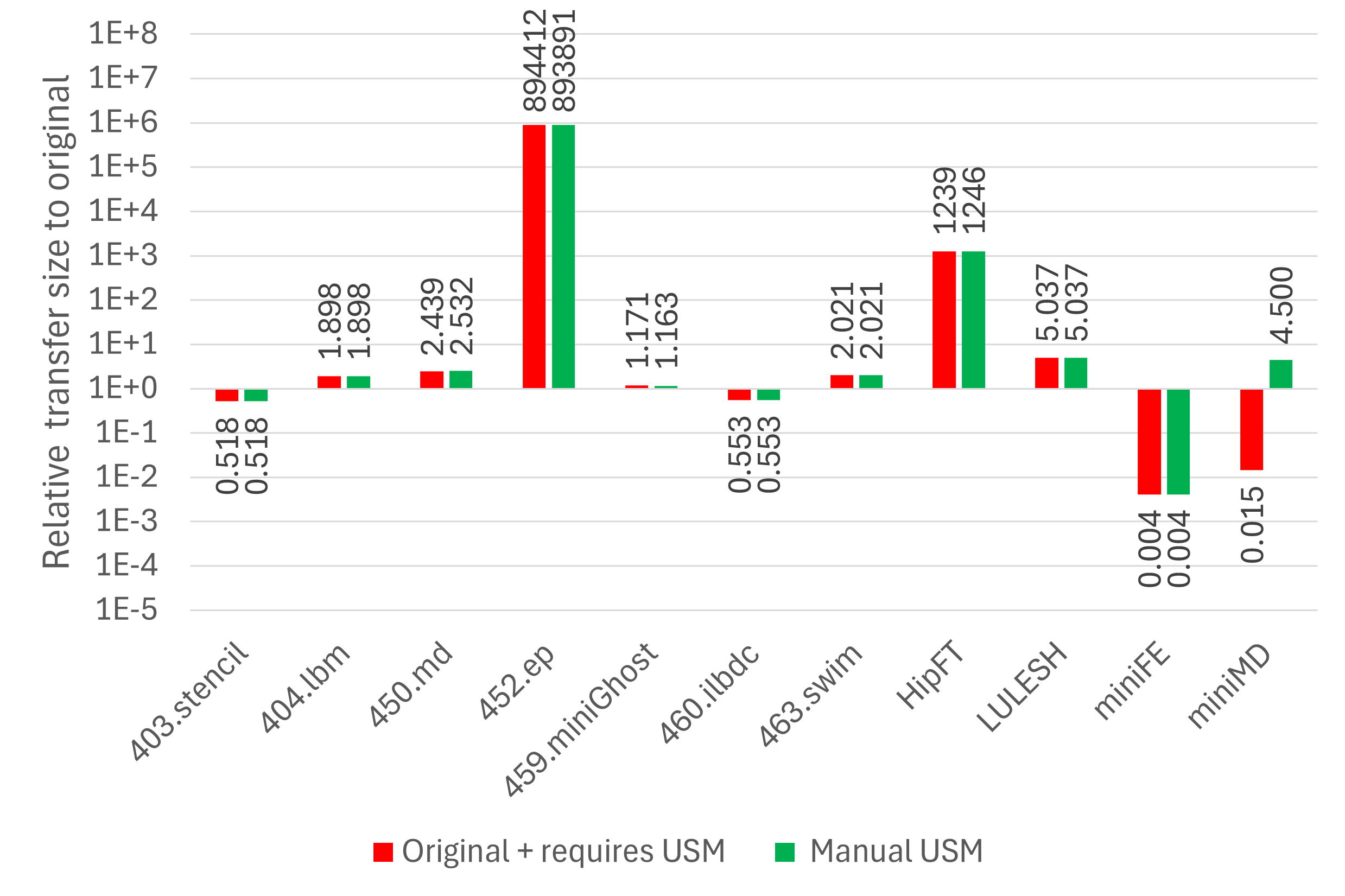}
        \caption{Relative transfer size of the USM versions when compared against the original version using explicit data transfers.}
        \label{fig:RelativeTransferSizeToOriginalVersion}
    \end{subfigure}
	\\
    \begin{subfigure}{0.485\textwidth}
        \includegraphics[scale=0.315]{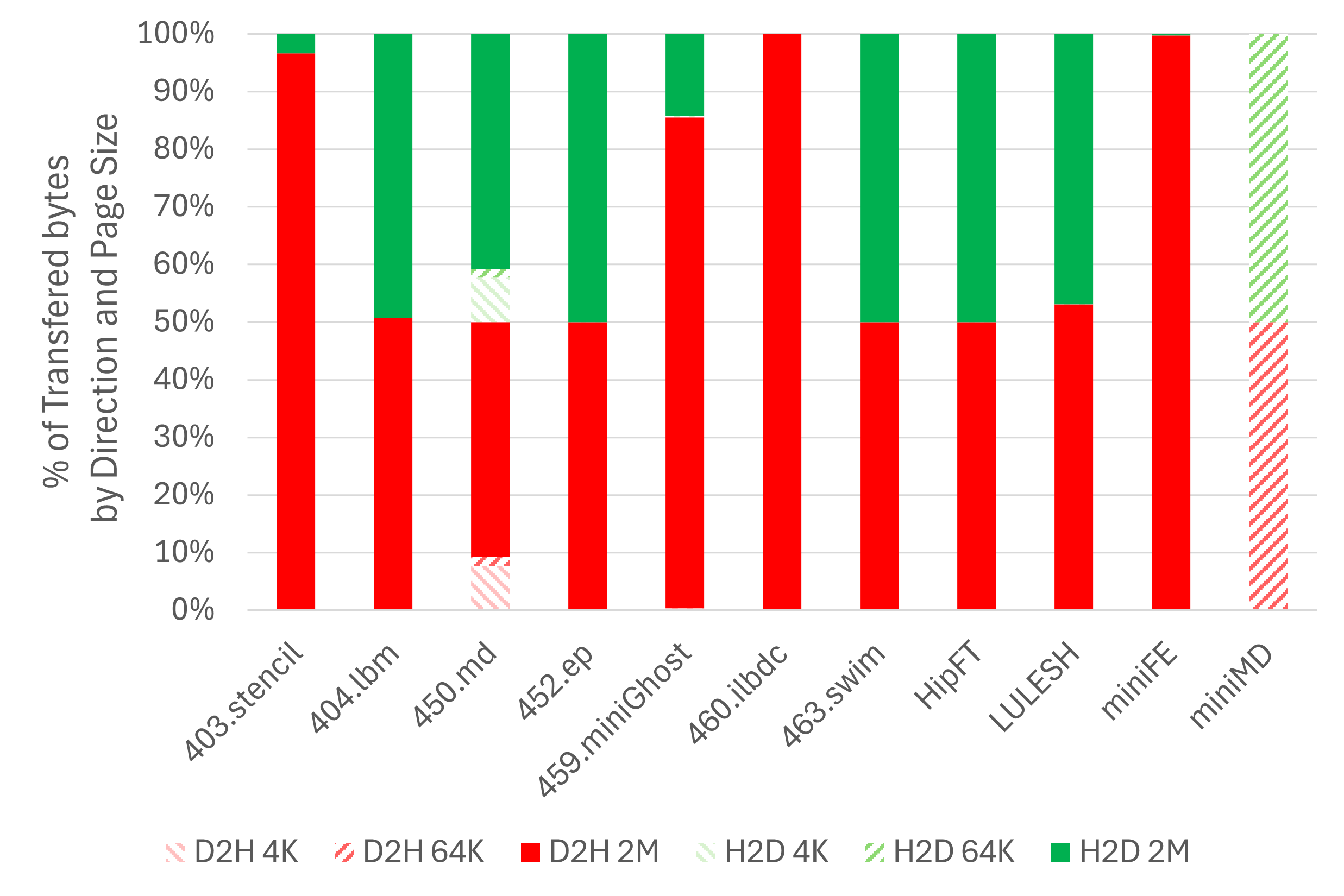}
        \caption{Migrations by page size for the \textit{automatic-USM} version.}
        \label{fig:TransfersByPageSize_ReqUSM}
    \end{subfigure}
	~
    \begin{subfigure}{0.485\textwidth}
        \includegraphics[scale=0.315]{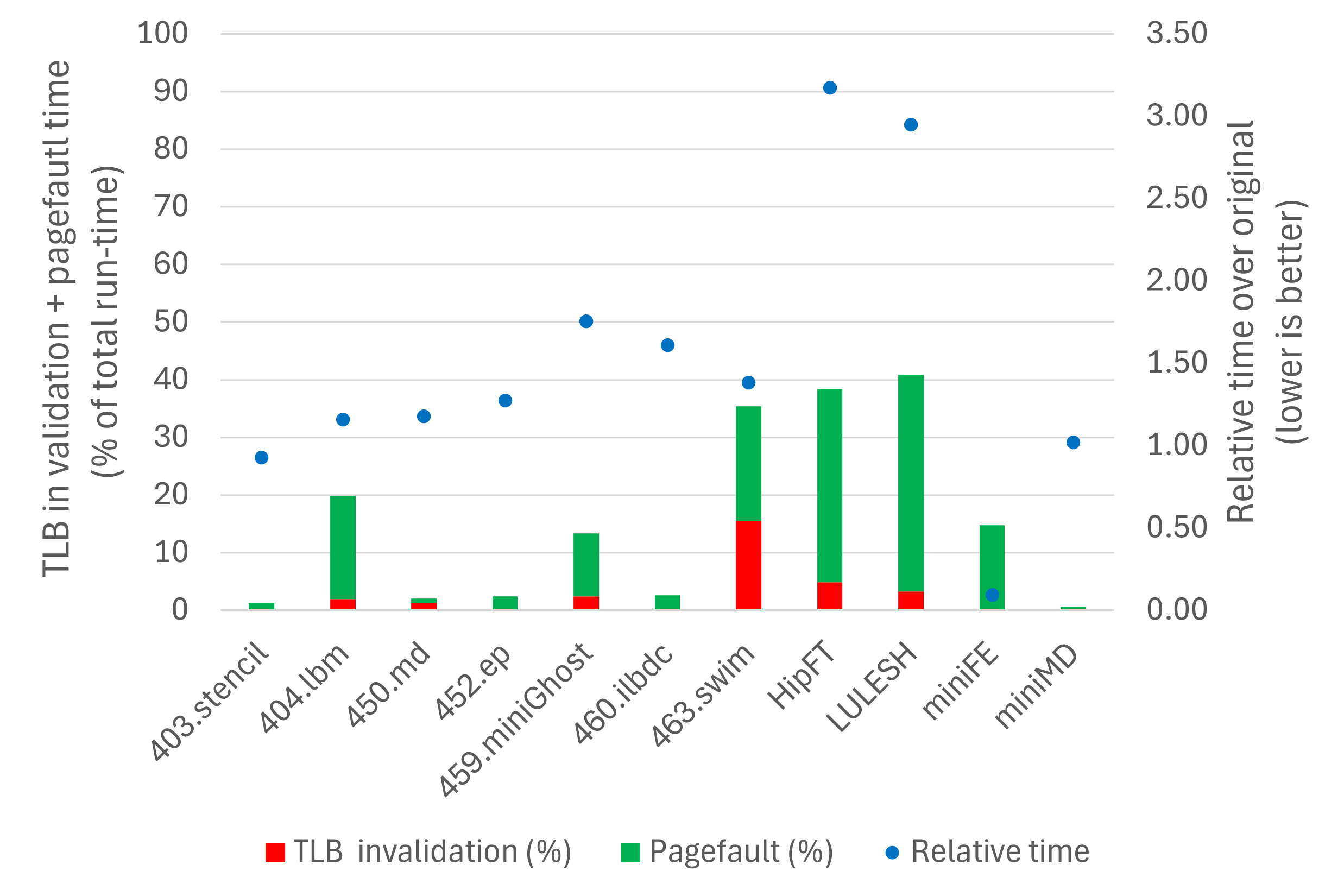}
        \caption{Page fault and TLB invalidation \% of time for the \textit{automatic-USM} version.}
        \label{fig:PageFaultAndTLBInvalidationRates_ReqUSM}
    \end{subfigure}
	\caption{Performance characteristics of the selected applications.}
\end{figure}

As mentioned earlier, one of the benefits of USM is the possibility of quick offload prototyping. For instance, when moving computation from the CPU to the GPU (assisted with USM) of LULESH and HipFT on this system, we see $2.73x$ and $1.64x$ performance gains, respectively. While this is a nice gain, in this section we evaluate how performance varies on applications already using OpenMP for accelerators and how USM impacts when it comes into play. Figure~\ref{fig:RelativePerformanceToOriginalVersion} shows the relative performance of the two USM versions when compared against the original code with explicit data transfers. USM versions exhibit a performance overhead with geometric mean below $1.2x$ and values ranging from $0.09x$ (miniFE) to $3.32x$ (HipFT). Two applications (miniFE and 403.stencil) show an improvement when using USM. Our first conclusion is that the \textit{automatic-USM} version can be easily tested on already existing OpenMP applications to explore potential performance benefits. The results are also surprising for 452.ep -- the embarrassingly parallel benchmark -- because it shows an overhead ($1.26x$) when using USM and we expect a performance on par with the original version.

There may be many factors contributing to the performance overhead of USM, including specific application characteristics (\textit{e.g.}, memory access patterns, parallelism, compute intensity, etc.). In this section, we focus the exploration around topics that may be attributable and measurable on the USM implementation, such as: 1) data transfer size, 2) page size employed at transfer time, and 3) observed page fault and TLB invalidation rates.

\subsubsection{Data transfer size}

The amount of data transferred between the host and the accelerator is an important factor to understand the performance overhead. When adopting USM, the data transfer size may increase or decrease when compared against the original version. For instance, it may increase if data objects involved in the computations are smaller than the page size because all the pages containing the data objects shall be transferred instead of just the data objects themselves. On the other hand, the transferred size may decrease if the data object is much larger than the working set and the page size. Intuitively, we expect a correlation between the transfer size ratio and the timing ratio.

Figure~\ref{fig:RelativeTransferSizeToOriginalVersion} shows the relative transfer size of the two USM versions when compared against the original version using explicit data transfers. The data transfers from the original version were collected through unitrace~\cite{Unitrace_URL}, while the data transfers from the USM versions were collected through the exposed OS metrics. Some results are aligned to the previous intuition, but the following are surprising, if not counter-intuitive:

\begin{itemize}
    \item 452.ep: shows a huge transfer size increase when using USM ($8.9\times10^5x$),
    \item 459.miniGhost: the performance overhead ($1.75x$) seems excessive for the modest transfer size increase ($1.17x$),
    \item 460.ilbdc: the application becomes slower ($1.63x$) despite the smaller transfer size ($0.55x$),
    \item HipFT: shows a huge transfer size increase when using USM ($1200x$) with high performance overhead ($3.17x$),
    \item LULESH: shows high performance overhead ($2.95x$) but the transfer size increase is not the largest, and
    \item miniMD: shows a huge diverging behavior between the two USM versions.
\end{itemize}

From above: 452.ep, LULESH, and HipFT suffer from a similar issue. All of them use \texttt{map(alloc)} with large arrays (especially 452.ep) or in nested loops (like HipFT). When using USM, the OpenMP runtime ignores the allocation and when the device runs the offloaded kernel the driver (see Section~\ref{sec:implementation-kmd}) starts pulling the memory from the host as soon as the kernel references the data object, even though it is unnecessary in circumstances where a \texttt{map(alloc)} is followed by an access to the corresponding list-item on device, without an intervening \texttt{target update to()} or \texttt{map(always,to/from)} to propagate the value of the original list-item to the corresponding list-item. Regarding 460.ilbdc, unitrace reports that when it comes to transfers, these are dominated by two big transfers (host to device at the beginning, device to host at the end), but collected numbers from the OS metrics seem to disregard one of the (page-fault assisted) transfers when using USM. We are still investigating this issue.
The miniFE code contains multiple data objects being transferred when they could reside on the GPU memory. The performance of the USM version is better, as expected, as it incurs smaller data transfers compared to the original, but the original code could be optimized by avoiding redundant data transfers on these objects. This shows how using USM can help identify potential for improvement in original applications that already have explicit data-motion controls.
Finally, the large difference between the two USM versions in miniMD is explained by the fact that the application uses \texttt{omp\_target\_associate\_ptr()} to associate host pointers with device pointers. When we adopted the \textit{manual-USM} approach, we inadvertently discarded these calls (and the associated \texttt{omp\_target\_alloc()}), but these calls still exist in the \textit{automatic-USM} version and as a result the transfers are done explicitly and not counted in the OS metrics. We have decided to keep them this way for the subsequent discussion.

\subsubsection{Migrations by page size}

The transfer efficiency depends on the size of the data being transferred, where typically, the larger the data size, the more efficient the transfer. Therefore, we explore the migrations by page size to understand the performance overhead of USM. Figure~\ref{fig:TransfersByPageSize_ReqUSM} shows the device to host (D2H) and host to device (H2D) migrations by page size for the \textit{automatic-USM} version. We observe that miniMD is mostly migrating the data using 64K pages. We also see in 450.md that about 10\% of D2H and another 10\% of H2D migrations use 4K pages. The remaining applications typically use 2MB pages to migrate data in either direction. This plot also shows that some applications (403.stencil, 459.miniGhost, 460.ilbdc, and miniFE) are dominated by D2H transfers.

\subsubsection{Page fault and TLB invalidation rates}

Page fault and TLB invalidation rates play a role to understand the performance overhead of USM, as discussed in Section~\ref{sec:implementation-kmd}. When using USM, page faults may occur when the compute kernel accesses a page that is not currently resident in the device memory, which can lead to significant performance degradation. To some extent, page faults are related to the transfer size (the more data transferred, the higher the likelihood of page faults) and page size used (the larger the page size, the lower the likelihood of page faults). In a related direction, TLB invalidations occur after the OS alters the CPU page table with an active GPU page mapping.

Figure~\ref{fig:PageFaultAndTLBInvalidationRates_ReqUSM} shows the percentage of time during which applications were servicing page faults and TLB invalidations on the \textit{automatic-USM} version (on the left Y-axis) by the means of bars while the dots represent the relative performance against the original version on the right Y-axis. The Figure shows a partial correlation, HipFT and LULESH spend about 40\% of their execution time servicing page faults and TLB invalidations, but it is also true that 404.lbm, 459.miniGhost and miniFE spend between 10 and 20\% of time servicing page faults and these applications do not experience a big performance penalty. 463.swim would be on the other side, it spends about 35\% of its time servicing page faults and TLB invalidations, but the performance is not as bad as experienced in LULESH nor HipFT. These results, while valuable, indicate that other topics in addition to page faults and TLB invalidations, like memory access patterns or the ability from the CPU or GPU threads to overlap their execution with the fault or invalidation servicing, are also playing a role in the performance overhead of USM.

\subsubsection{Improving USM performance}

We have seen that some applications spend time to transfer data from the host to the device even when the data is uninitialized. To avoid such expense, and while modifying the sources is beyond the purpose of this paper, we have experimented using the \texttt{close} modifier on the \texttt{map(alloc)} clause on applications using it (namely: 404.lbm, 450.md, 452.ep, 463.swim, HipFT and LULESH). All applications completed successfully and validated the results except for 404.lbm -- we are investigating this failure.

Those applications that completed successfully show an interesting performance improvement (plots not added due to space restrictions). These new results show: 1) 463.swim runs faster than the original version, 2) a $2.1x$ performance improvement on HipFT when compared to the \textit{automatic-USM} version, 3) 452.ep running on par compared with the original version, and 4) other modest improvements on LULESH and 450.md. Based on the metrics captured during the runs, these improvements can be attributed to the significant reductions in the transfer size and in page fault rates. If we take this improvement into account, the geometric mean of the USM overhead drops to $1.03x$ when compared against the original version with explicit transfers.

\section{Further Discussion}
\label{sec:FurtherDiscussion}

When it comes to programmability, although we have shown that it is non-intrusive to adopt the \textit{automatic-USM} approach, this might be more involved for huge applications with many compilation units, included files or many Fortran modules and/or evolving codes, especially if developers do not want to blindly add the directive. In a similar way to NVIDIA's HPC SDK compiler suite\footnote{Through the \texttt{-gpu=mem:unified} compiler flag.}, the Intel\textsuperscript{\textregistered} oneAPI DPC++/C++ Compiler has a command-line flag\footnote{\texttt{-fopenmp-force-usm}} that avoids the need for explicitly adding the \texttt{requires} directive. A similar command-line flag will be implemented in subsequent versions of the Intel Fortran Compiler, helping with the adoption of USM and also helping applications that adopted \texttt{do concurrent} to be fully agnostic to explicit programming models.

Regarding performance, there are a number of topics worth mentioning. First, there are a number of Xe kernel driver optimizations in the pipeline aimed at reducing the GPU page-fault flooding that may help address some of the issues described earlier, but we cannot determine when these optimizations will be publicly available. Second, the compiler team is looking into ways to further improve performance through hints to Level Zero to specify residency of arrays that may achieve similar gains without the need of the \texttt{close} modifier. Third, it may be worth exploring how to efficiently map \texttt{omp\_target\_associate\_ptr} and \texttt{omp\_target\_disassociate\_ptr} into the USM implementation, where an initial approach could be the KMD to honor the OpenMP requested association. And fourth, the Intel compiler implements an extension to the OpenMP specification for prefetching data\footnote{\url{https://www.intel.com/content/www/us/en/docs/dpcpp-cpp-compiler/developer-guide-reference/2025-2/ompx-prefetch-data.html}} from the accelerator. It may be worth extending it to prefetch data objects across devices.

Finally, USM opens the door to additional performance analyses. These cover understanding which data objects are used on each device and how, learning what the USM mechanism does in the background, correlating this information with application sources, \textit{etc}. Related to this, USM has shown that optimizing map clauses and their array sections may lead to important performance uplifts.

\section{Conclusions}
\label{sec:Conclusions}

USM is a great addition to the OpenMP specification, allowing programmers to easily write and maintain code for heterogeneous systems. In this paper, we have described the implementation, explained how to program with USM, and provided some initial performance results when adopting USM in a number of OpenMP applications on Intel accelerators. When comparing already existing OpenMP applications and incorporating USM into them, we have seen a performance overhead of about $1.2x$ (which was brought down to $1.03x$ after manual source changes), which makes us confident that USM is a promising feature for OpenMP programmers. Moreover, given that some of these applications observed a performance uplift and that USM is easy to adopt, we encourage programmers to give this feature a try and see if it can help them to achieve better performance in their applications. Finally, we have discussed potential topics of interest around USM, ranging from new compiler flags, to kernel optimizations, OpenMP extensions and ideas for performance analysis.

\subsection*{Acknowledgements}

We want to thank Ron Caplan for his constructive comments and contributions to this research paper.

\subsection*{Notices \& Disclaimers}
\small{
Performance varies by use, configuration and other factors.
Performance results are based on testing as of dates shown in configurations and may not reflect all publicly available updates.
No product or component can be absolutely secure.
Intel disclaims all express and implied warranties, including without limitation, the implied warranties of merchantability, fitness for a particular purpose, and non-infringement, as well as any warranty arising from course of performance, course of dealing, or usage in trade.
Your costs and results may vary.
Intel does not control or audit third-party data. You should consult other sources to evaluate accuracy.
Intel technologies may require enabled hardware, software or service activation.

\textsuperscript{\textcopyright}Intel Corporation. Intel, the Intel logo, and other Intel marks are trademarks of Intel Corporation or its subsidiaries.

* Other names and brands may be claimed as the property of others.
}

%
%
%
\bibliographystyle{splncs04}
\bibliography{bibliography-compact}

\end{document}